\documentclass[final,5p,twocolumn,times]{elsarticle}

\usepackage{amssymb}

\journal{Energy Strategy Reviews}

\newcommand{\fref}[1]{Figure~\ref{#1}}

\newcommand{\tref}[1]{Table~\ref{#1}}
\newcommand{\highlight}[1]{{\color{black}#1}}

\usepackage{arydshln}

\usepackage[colorlinks]{hyperref}

\usepackage[final]{pdfpages}

\begin{document}

\begin{frontmatter}



\title{Real-Time Carbon Accounting Method for the European Electricity Markets}


\author[label1,label2,label3]{Bo Tranberg\corref{cor1}}
\author[label4]{Olivier Corradi}
\author[label4]{Bruno Lajoie}
\author[label5]{Thomas Gibon}
\author[label6]{Iain Staffell}
\author[label2]{Gorm Bruun Andresen}

\address[label1]{Ento Labs ApS, Inge Lehmanns Gade 10, 6., 8000 Aarhus C, Denmark}
\address[label2]{Department of Engineering, Aarhus University, Inge Lehmanns Gade 10, 8000 Aarhus C, Denmark}
\address[label3]{Danske Commodities, V{\ae}rkmestergade 3, 8000 Aarhus C, Denmark}
\address[label4]{Tomorrow, TMROW IVS, \href{https://www.tmrow.com}{tmrow.com}, Godth{\aa}bsvej 61 B, 3. th., 2000 Frederiksberg}
\address[label5]{Luxembourg Institute of Science and Technology, 5 Avenue des Hauts-Fourneaux, 4362 Esch-sur-Alzette, Luxembourg}
\address[label6]{Centre for Environmental Policy, Imperial College London, London, UK}
\cortext[cor1]{Corresponding author: bo@entolabs.co}

\date{\today}

\begin{abstract}
Electricity accounts for 25\% of global greenhouse gas emissions. Reducing emissions related to electricity consumption requires accurate measurements readily available to consumers, regulators and investors. In this case study, we propose a new real-time consumption-based accounting approach based on flow tracing. This method traces power flows from producer to consumer thereby representing the underlying physics of the electricity system, in contrast to the traditional input-output models of carbon accounting. With this method we explore the hourly structure of electricity trade across Europe in 2017, and find substantial differences between production and consumption intensities. This emphasizes the importance of considering cross-border flows for increased transparency regarding carbon emission accounting of electricity.
\end{abstract}

\begin{keyword}
carbon accounting \sep carbon emission \sep carbon intensity \sep flow tracing


\end{keyword}

\end{frontmatter}


\section{Introduction}\label{sec:intro}
For several decades, more than 80\% of the global electricity generation has been generated from fossil fuel \cite{iea2018b}. As a result, electricity and heat production account for 25\% of global greenhouse gas (GHG) emissions \cite{IPCCreport}. Furthermore, electricity demand is widely expected to rise because of electrification of vehicles \cite{iea2018}. These facts highlight the importance of an accurate and transparent carbon emission accounting system for electricity.

Reducing emissions related to electricity consumption requires accurate measurements readily available to consumers, regulators and investors \cite{Brander2018}. In the GHG protocol \cite{WRI}, ``Scope 2 denotes the point-of-generation emissions from purchased electricity (or other forms of energy)'' \cite{Brander2018}. A major challenge regarding Scope 2 emissions is the fact that it is not possible to trace electricity from a specific generator to a specific consumer \cite{Jiusto2006,Raadal2013}. This has lead to the use of two different accounting methods: the of \emph{grid average} emission factors or the \emph{market-based} method \cite{Brander2018,Raadal2013}. Grid average factors are averaged over time and therefore not specific to the time of consumption due to limited availability of emission factors with high temporal resolution. The market based method entails purchasing contractual emission factors in the form of different types of certificates, which do not affect the amount of renewable electricity being generated, and therefore fail to provide accurate information in GHG reports. For a detailed criticism of both approaches, see \cite{Brander2018}.

In this case study, we propose a new method for real-time carbon accounting based on flow tracing techniques. This method is applied to hourly market data for 28 areas within Europe. We use this method to introduce a new consumption-based accounting method that represents the underlying physics of the electricity system in contrast to the traditional input-output models of carbon accounting \cite{Fan2016,Zhang2018,Clauss2018}. The approach advances beyond \cite{carbon-ft}, where a similar flow tracing methodology is used to create a consumption-based carbon allocation between six Chinese regions. However, the data for that study was limited to annual aggregates and different generation technologies were also aggregated. We apply the method to real-time system data, including the possibility of distinguishing between different generation technologies, \highlight{providing a real-time CO\textsubscript{2} signal for all actors involved.} This increases the overall transparency and credibility of emission accounting related to electricity consumption, which is of high importance \cite{Sjodin2004}. To investigate the impact of the new consumption-based accounting method we compare it with the straightforward production-based method (i.e. looking at the real-time generation mix within each area). For discussions on the shift from production-based to consumption-based accounting and the idea of sharing the responsibility between producer and consumer, we refer to \cite{Lenzen2007,Peters2008}.

\section{Methods}\label{sec:methods}
\subsection{Data}
\begin{figure}[t]
\centering
\includegraphics[width=\columnwidth]{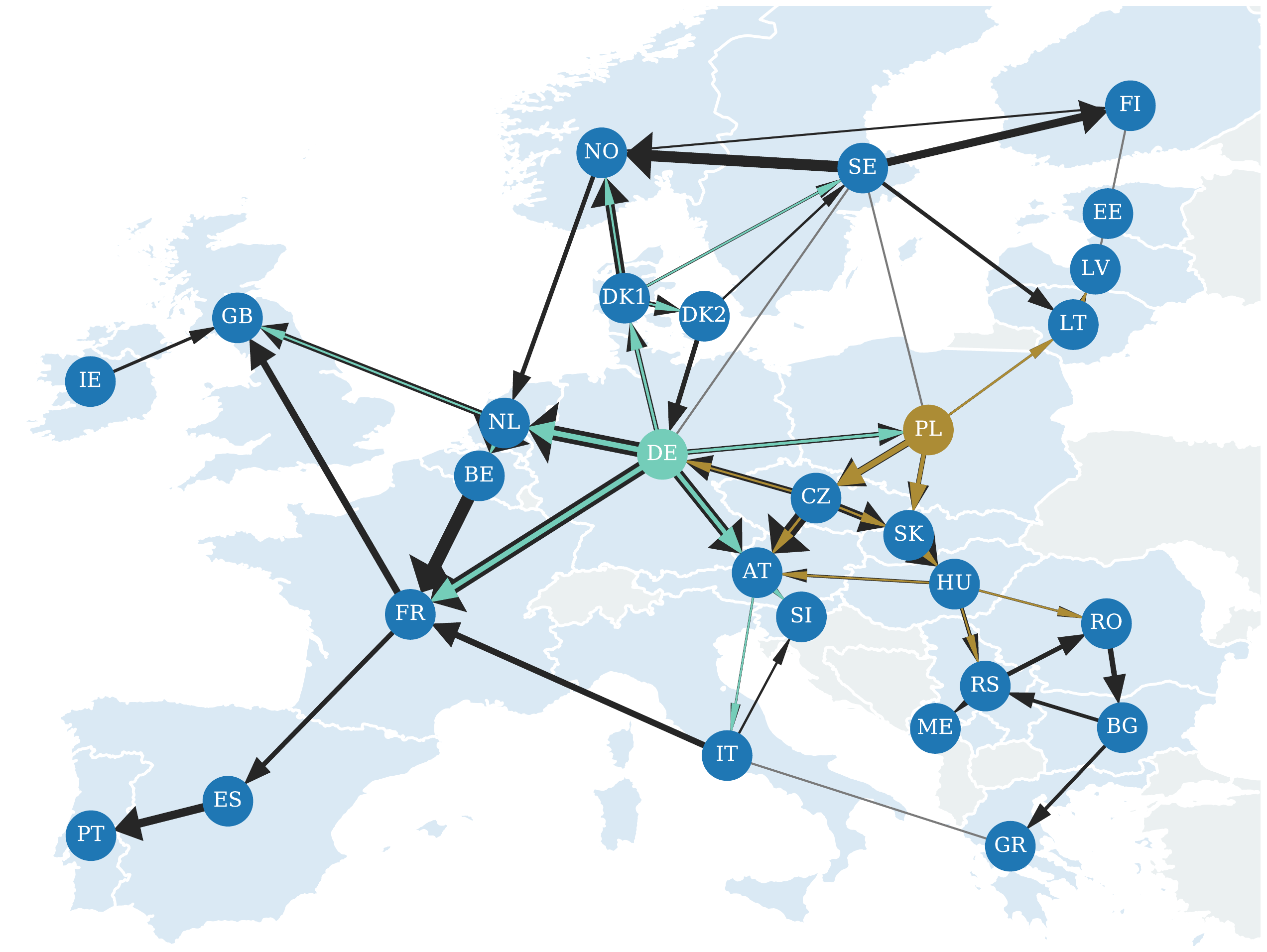}
\caption{The 28 areas considered in this case study, and the power flows between them for the first hour of January 1, 2017. The width of the arrows is proportional to the magnitude of the flow on each line. Power flows to and from neighboring countries, e.g. Switzerland, are included when available, and these areas are shown in gray. The cascade of power flows from German wind and Polish coal are highlighted with blue and brown arrows, respectively.}
\label{fig:topology}
\end{figure}

The method is applied to data from the electricityMap database \cite{electricityMapDatabase}, which collects real-time data from electricity generation and imports/exports around the world. The European dataset, consisting of 28 areas, is used with hourly resolution for the year 2017. Data sources for each individual area can be found on the project's webpage \cite{electricityMap}. \fref{fig:topology} shows the 28 areas and the 47 interconnectors considered. Power flows to and from neighboring areas, e.g. Switzerland, are included when available. The black arrows show a snapshot of hourly power flows between the areas. In the results, we aggregate the two price areas of Denmark and, thus, compare 27 countries.

\begin{figure}[t]
\centering
\includegraphics[width=\columnwidth]{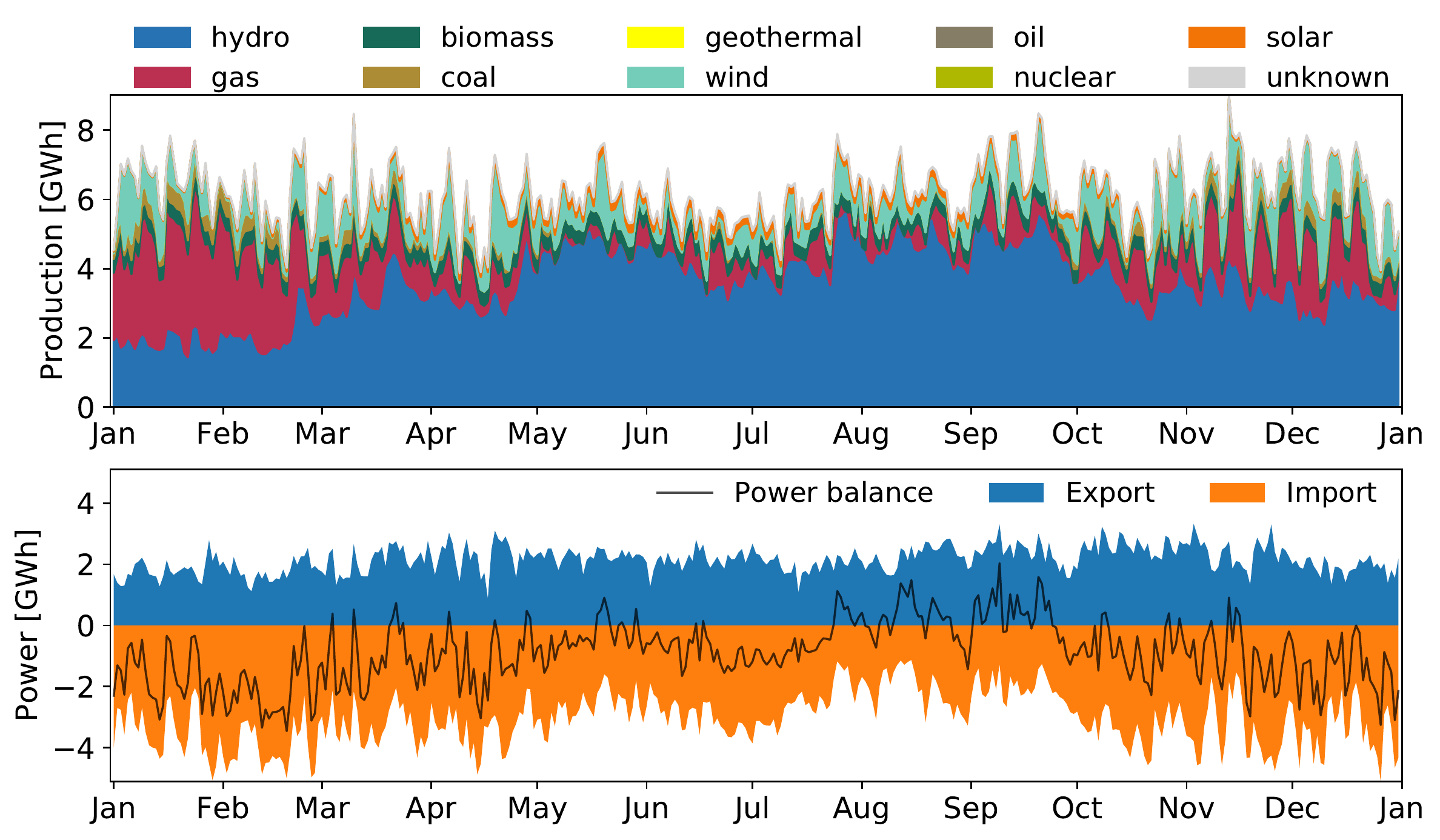}
\caption{Daily-average stacked power production for each technology for Austria during 2017 (top) as well as exports, imports and power balance (bottom).}
\label{fig:production}
\end{figure}

The top panel of \fref{fig:production} shows stacked daily-average production for each technology for Austria. The bottom panel shows daily-average exports and imports. The black line represents the sum of the hourly exports and imports showing Austria's net import/export position. The daily averages in this figure are based on the full 8760 hours in the dataset representing the full year 2017.

\begin{table}[t]
\caption{CO\textsubscript{2} equivalent operation intensity per technology averaged across countries. The dashed line indicates the split between non-fossil and fossil technologies. For details, see Table~1--3 in the supplementary material.}
\label{tab:co2}
\centering
\small
\begin{tabular}{ll}
\hline
Technology & Intensity [kgCO\textsubscript{2}eq/MWh]\\
\hline
solar              &  \hspace{14.5mm}0.00410\\
geothermal         &  \hspace{14.5mm}0.00664\\
wind               &  \hspace{14.5mm}0.141\\
nuclear            &  \hspace{12.5mm}10.3\\
hydro			   &  \hspace{12.5mm}16.2\\
biomass            &  \hspace{12.5mm}50.9\\
\hdashline
gas                &  \hspace{11mm}583\\
unknown            &  \hspace{11mm}927\\
oil                &  \hspace{9mm}1033\\
coal               &  \hspace{9mm}1167\\
\hline
\end{tabular}
\end{table}

Carbon emission intensities are derived from the ecoinvent 3.4 database to construct an accurate average intensity per generation technology per country decomposed in lifecycle, infrastructure and operations \cite{ecoinvent}. The operations intensities are used for the production and consumption-based carbon allocation in this study. Operational emissions include all emissions occurring over the fuel chain (from extraction to supply at plant) as well as direct emissions on site. For fossil fuels, operational emissions are therefore higher than only direct combustion emissions. For solar, geothermal and wind, the emissions are strictly from maintenance operations.

The operations intensity per technology averaged over all countries is summarized in \tref{tab:co2}. The dashed line indicates the split between non-fossil and fossil technologies. For details on country-specific values, see Table~1--3 in the supplementary material.

\subsection{Carbon emission allocation}
The consumption-based accounting method proposed in this case study builds on flow tracing techniques. Flow tracing was originally introduced as a method for transmission loss allocation and grid usage fees \cite{Bialek,Kirschen1997}. It follows power flows on the transmission network mapping the paths between the location of generation and the location of consumption. It works in such a way that each technology for each country is assigned a unique color mathematically. This is a mathematical abstraction since it is not physically possible to color power flows. For each hour local production and imported flows are assumed to mix evenly at each node in the transmission network (see \fref{fig:topology}) and determine the color mix of the power serving the demand and the exported flows. As an example, the colored arrows in \fref{fig:topology} show the cascade of power flows resulting from flow tracing of German wind power (light blue) and Polish coal power (brown) for the first hour of January 1st, 2017. The size of the colored arrows shows how much of the total power flow (in black) is accounted for. A threshold has been applied such that the technology specific flows are only shown if they account for at least 2\% of the total power flow for each interconnector.

Flow tracing has been proposed as the method for flow allocation in the Inter-Transmission System Operator Compensation mechanism for transit flows \cite{itc,Acer2013}. Recently, the method has been applied to various aspects of power system models to allocate transmission network usage \cite{Tranberg2015,Schaefer2017}, a generalization that allows associating power flows on the grid to specific regions or generation technologies \cite{Hoersch2018}, creating a flow-based nodal levelized cost of electricity \cite{Tranberg2018a}, and analyzing the usage of different storage technologies \cite{Tranberg2018b}.

The challenge of cross-border power flows in relation to carbon emission accounting has previously been studied in \cite{Jiusto2006,carbon-ft}. Both studies simplify nodes as being either net importers or net exporters and neither are able to distinguish between different generation technologies. Those simplifications are not necessary in our approach as we can deal with both imports, exports, consumption and generation simultaneously at every node while also distinguishing between different generation technologies. Additionally, \fref{fig:topology} exhibits loop flows. However, these do not affect the validity of the flow tracing methodology \cite{carbon-ft}, and no effort has been made to eliminate them as they occur naturally in the transmission system at the area level \cite{Kunz2018}.

Flow tracing methods are almost unanimously applied to simulation data -- typically with high shares of renewable energy. In this case study, we apply the flow tracing method to hourly time series from the electricityMap \cite{electricityMap}. From this we are able to map the power flows between exporting and importing countries for each type of generation technology for every hour of the time series. Applying country-specific average carbon emission intensity per generation technology to this mapping, we construct a consumption-based carbon accounting method. For details on the mathematical definitions, see Section~B in the supplementary material.

The production-based accounting method used for comparison, is calculated as the carbon intensity from local generation within each country.

\begin{figure}[t]
\centering
\includegraphics[width=\columnwidth]{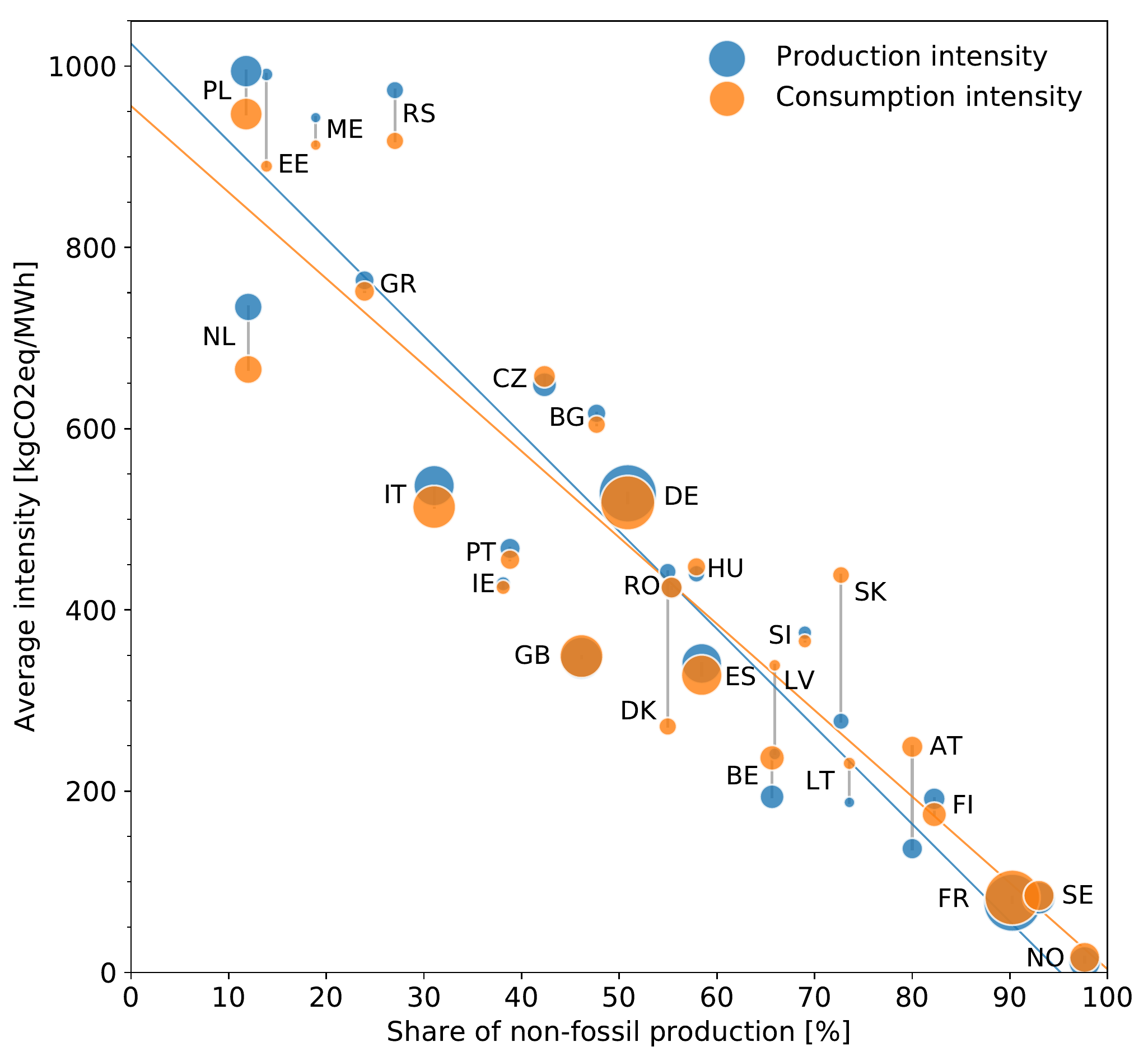}
\caption{Comparison of average hourly production and consumption intensity as a function of the share of non-fossil generation in the country's generation mix. Size of circles are proportional to mean generation and mean consumption for each country.}
\label{fig:scatter}
\end{figure}

\section{Results}\label{sec:results}
\fref{fig:scatter} shows a comparison of average production and consumption intensity as a function of the share of non-fossil generation in each country's generation mix. The consumption intensity is calculated using flow tracing. The size of the circles is proportional to the average hourly generation and consumption in MWh, respectively. A vertical gray line connects the production and consumption intensity corresponding to the same country. We see a decline in intensity with increasing share of non-fossil generation. For high shares of non-fossil generation, the consumption intensity tends to be higher than the production intensity due to imports from countries with higher production intensity. The pattern is reversed for low shares of non-fossil generation. The values plotted in this figure are shown in Table~4 in the supplementary material.

Some countries exhibit a huge difference between production and consumption intensity. An example of this is Slovakia (SK), which has a high share of nuclear power and Austria (AT), which has a high share of hydro power, but both rely heavily on imports of large amounts of coal power especially from Poland (PL) and Czech Republic (CZ). Denmark (DK) is an extreme example of the opposite case, having a high share of coal and gas power and importing large amounts of hydro and nuclear power from Norway (NO) and Sweden (SE).

While this figure only shows average values, Figure~7 in the supplementary material highlights the interval of hourly variation of production and consumption intensity per country. This interval is high for all countries except the ones with very high non-fossil share (FR, SE, NO).

From a national perspective, it is important to know the source electricity that is being imported, and whether it increases a country's reliance on high-carbon, insecure, or otherwise undesirable sources of generation.

\begin{figure}[t]
\centering
\includegraphics[width=\columnwidth]{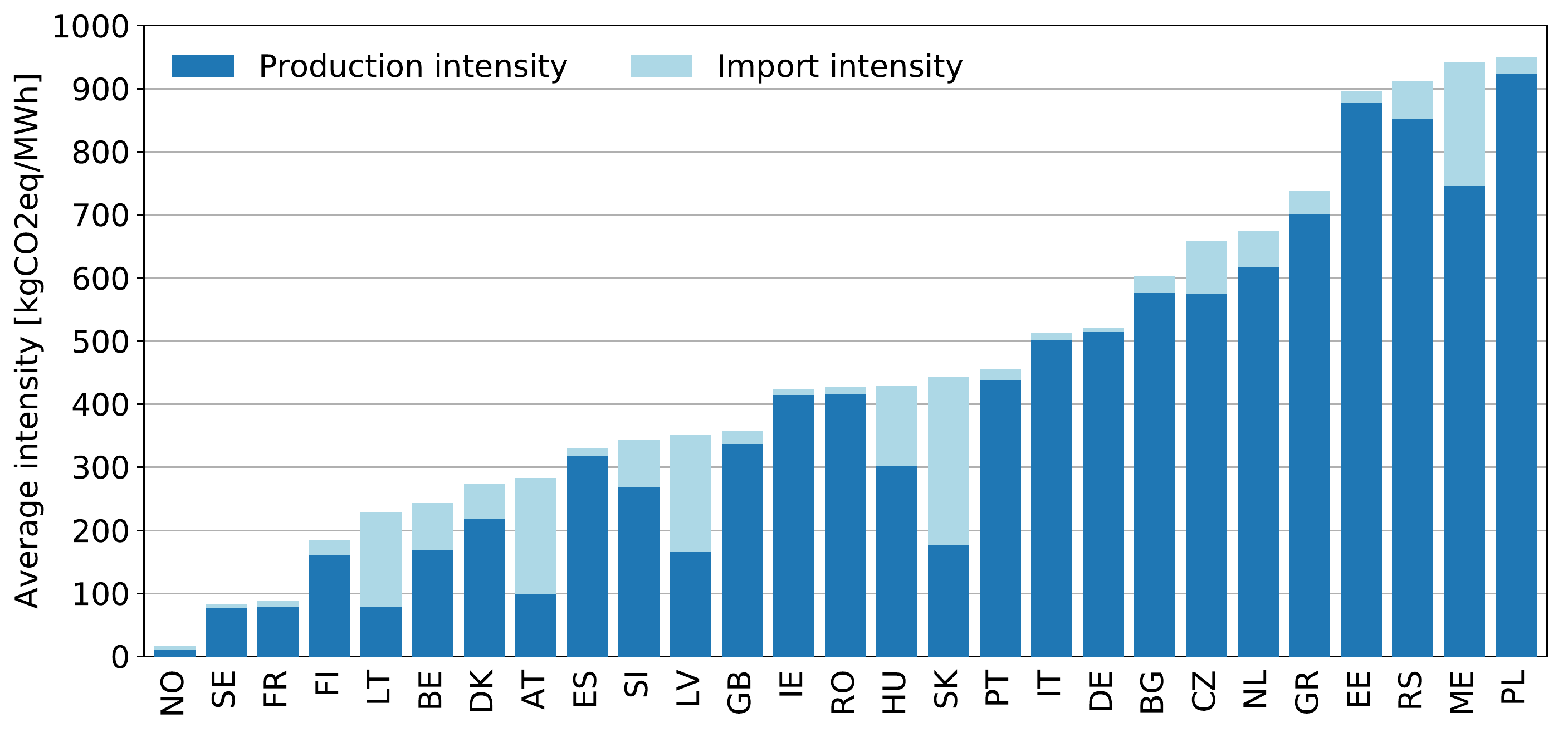}
\caption{Average hourly consumption intensity per consumed MWh per country (stacked bar) split in contributions from local generation and imports. The countries are sorted by average consumption intensity.}
\label{fig:prod-import-intensity}
\end{figure}

\fref{fig:prod-import-intensity} shows the consumption-based intensity per country. The height of each bar corresponds to the consumption intensity for each country shown in \fref{fig:scatter}. This figure decomposes the consumption intensity for each country and shows how much of a particular country's consumption intensity is caused by the local generation mix compared with the generation mix of imported power. We see that for many countries it is important to be able to distinguish between local generation and imports since the imports make a substantial contribution to the country's consumption-based emission. In cases with a large difference between the intensity of local power production and the imported power, imports have a high impact. As mentioned in an earlier example, this is the case for both Austria and Slovakia. For  details on the average intensity of imports and exports between the countries, see Figure~9 and Table~5 in the supplementary material.

\section{Conclusion}\label{sec:conclusion}
\highlight{
We introduce a new method for consumption-based carbon emission allocation based on flow tracing applied to a historical sample of real-time system data from the electricityMap.

The method we propose demonstrates that consumption-based accounting is more difficult than production-based due to the added complexity of cross-border flows. However, with this method we have found substantial differences between production and consumption intensities for each country considered, which follow a trend proportional to the share of non-fossil generation technologies. It would be straightforward to subsequently apply these results to attribute carbon emissions to individual consumers like companies or households.

The difference between production and consumption intensities and the associated impact of imports on average consumption intensity emphasize the importance of including cross-border flows for increased transparency regarding carbon emission accounting of electricity. While there are limitations to the accuracy of this method due to data availability and the mathematical abstraction of flow tracing, we believe that this method provides the first step in a new direction for carbon emission accounting of electricity.

This case study focuses on the European electricity system. When additional sources of live system data become available this approach could be extended to cover a wider geographical area. Even for areas without significant import and export the method could be applied within a single country provided that local system data is available at high spatial resolution. Another interesting application of this method would be to include additional sectors such as heating and transport as these are becoming electrified. This could lead to a real-time carbon emission signal for the entire energy system and potentially lay the foundation for time-varying electricity taxes.
}

\section*{Acknowledgments}
Gorm Bruun Andresen acknowledges the APPLAUS project for financial support. Iain Staffell acknowledges the Engineering and Physical Sciences Research Council (EPSRC) for funding via project EP/N005996/1. We thank Mirko Sch{\"a}fer for helpful discussions.



\section*{References}
\bibliographystyle{elsarticle-num}
\bibliography{references}








\section*{Supplementary material}

# Supplementary material to: Real-Time Carbon Accounting Method for the European Electricity Markets

Bo Tranberg[a,b,c], Olivier Corradi[d], Bruno Lajoie[d], Thomas Gibon[e], Iain Staffell[f] and Gorm Bruun Andresen[b]

[a]Ento Labs ApS, Inge Lehmanns Gade 10, 6., 8000 Aarhus C, Denmark
[b]Department of Engineering, Aarhus University, Inge Lehmanns Gade 10, 8000 Aarhus C, Denmark
[c]Danske Commodities, Værkmestergade 3, 8000 Aarhus C, Denmark
[d]Tomorrow, TMROW IVS, tmrow.com, Godthåbsvej 61 B, 3. th., 2000 Frederiksberg
[e]Luxembourg Institute of Science and Technology, 5 Avenue des Hauts-Fourneaux, 4362 Esch-sur-Alzette, Luxembourg
[f]Centre for Environmental Policy, Imperial College London, London, UK

May 15, 2019

## Contents

# A  Carbon intensities

Carbon emission intensities are derived from the ecoinvent 3.4 database [1]. For each of the EU28 we calculate technology-specific factors extracted from the high-voltage level (for most technologies) and low-voltage level (for photovoltaic technologies), to generate their lifecycle carbon intensities in grams of $CO_2$ equivalents per kilowatthour. Furthermore, we also differentiate infrastructure-related impacts from operational impacts. This is done by grouping life cycle inventory inputs by unit, where the set {'meter', 'meter-year', 'unit', 'kilometer'} are assumed to denote infrastructure processes, whereas the rest, that is, 'kilowatthour', 'tonne-kilometer', etc., are accounted as operation and maintenance processes.

The values under "high-voltage mix" denote the global warming potential (GWP) score of the electricity mix directly from high-voltage technologies, while "low-voltage mix" values denote the GWP score of electricity at the consumer level, i.e. after transformation and distribution from high and medium-voltage (including losses), and integration of photovoltaic electricity into the grid. The high- and low-voltage GWP scores are extracted directly from ecoinvent 3.4, here only shown for information, and never used in the calculations.

Not all technology-area pairs are available in the database, in case of missing information, values have been proxied by the EU28 average intensity for the given technology, calculated from the areas for which the data exists, and weighted by their respective contribution to the EU28 mix. When the production source is unknown we assume an intensity averaged over the particular country's intensity for gas, oil and coal.

Table 1–3 show the country-specific lifecycle, infrastructure, and operation intensities per technology in units of g $CO_2$ eq./kWh. EU28 averages are also shown, in bold. The relation between the three tables is such that lifecycle = infrastructure + operation. The operation intensities in Table 3 are the basis for the production as well as consumption-based carbon allocation in this study.

Table 1: Lifecycle $CO_2$ equivalent intensity per technology and country, in g $CO_2$ eq./kWh. Values in italic indicate that the country-specific factor is not available, and was replaced by the European weighted average for that technology (shown in bold).

| category | variant | AT | BE | BG | CZ | DE | DK | EE | ES | **EU28** | FI | FR | GB | GR | HU | IE | IT | LT | LV | ME | NL | NO | PL | PT | RO | RS | SE | SI | SK |
|---|---|---|---|---|---|---|---|---|---|---|---|---|---|---|---|---|---|---|---|---|---|---|---|---|---|---|---|---|---|
| high-voltage mix | - | 125 | 188 | 609 | 731 | 654 | 432 | 1030 | 336 | **426** | 262 | 41.9 | 801 | 980 | 400 | 513 | 469 | 551 | 520 | 426 | 616 | 15.9 | 1000 | 360 | 398 | 852 | 21.8 | 434 | 216 |
| wind | - | 17.8 | 16.2 | 19.5 | 19.4 | 20.0 | 13.8 | 19.8 | 14.2 | **16.8** | 23.0 | 15.6 | 16.8 | 15.1 | 13.6 | 13.7 | 19.7 | 13.3 | 18.2 | *16.8* | 16.3 | 14.4 | 16.5 | 13.7 | 25.3 | *16.8* | 16.2 | *16.8* | *16.8* |
| nuclear | - | *12.4* | 12.0 | 12.0 | 12.0 | 11.3 | *12.4* | *12.4* | 12.1 | **12.4** | 12.5 | 12.9 | *12.4* | *12.4* | 12.0 | *12.4* | *12.4* | *12.4* | *12.4* | *12.4* | 12.0 | *12.4* | *12.4* | *12.4* | 14.2 | *12.4* | 12.2 | 12.0 | 12.0 |
| geothermal | - | *81.8* | *81.8* | *81.8* | *81.8* | *81.8* | *81.8* | *81.8* | *81.8* | **81.8** | *81.8* | *81.8* | *81.8* | *81.8* | *81.8* | *81.8* | *81.8* | *81.8* | *81.8* | *81.8* | *81.8* | *81.8* | *81.8* | *81.8* | *81.8* | *81.8* | *81.8* | *81.8* | *81.8* |
| biomass | cogeneration | 53.8 | 53.8 | 56.8 | 53.8 | 53.8 | 53.8 | 56.8 | 53.8 | **53.9** | 53.8 | 53.8 | 53.8 | 53.8 | 53.8 | 53.8 | 53.8 | 56.8 | 56.8 | 53.9 | 53.8 | 53.9 | 53.8 | 53.8 | 53.8 | 53.9 | 53.8 | 53.8 | 53.8 |
| hydropower | pumped storage | 452 | 378 | 901 | 1140 | 965 | *617* | *617* | 546 | **617** | *617* | 77.3 | *617* | 1420 | *617* | 851 | 615 | 1040 | *617* | *617* | *617* | 41.7 | 1420 | 588 | 629 | 1220 | *617* | *617* | 684 |
|  | reservoir | 6.97 | *14.7* | *14.7* | *14.7* | 51.4 | *14.7* | *14.7* | 51.4 | **14.7** | 51.4 | 6.97 | *14.7* | *14.7* | *14.7* | *14.7* | 6.97 | *14.7* | *14.7* | *14.7* | *14.7* | 6.97 | *14.7* | 51.4 | *14.7* | 6.97 | 51.4 | *14.7* | 51.4 |
|  | run-of-river | 4.42 | 4.42 | 4.42 | 4.42 | 4.42 | 4.42 | 4.42 | 4.42 | **4.42** | 4.42 | 4.42 | 4.42 | 4.42 | 4.42 | 4.42 | 4.42 | 4.42 | 4.42 | 4.42 | 4.42 | 4.42 | 4.42 | 4.42 | 4.42 | 4.42 | 4.42 | 4.42 | 4.42 |
| coal | - | 986 | 1120 | 1180 | 1190 | 1170 | *1160* | 1300 | 1210 | **1160** | 1080 | 1090 | 1140 | 1300 | 1410 | 1070 | 1150 | 1180 | *1160* | *1160* | 1030 | *1160* | *1160* | 1140 | 1140 | 1340 | 1180 | 1200 | *1160* |
|  | cogeneration | 1220 | *1210* | 1250 | 1710 | 1170 | 1050 | *1210* | *1210* | **1210** | 1100 | *1210* | *1210* | 1560 | 1240 | *1210* | 1260 | *1210* | *1210* | *1210* | 998 | 1490 | 1160 | *1210* | 1240 | 1240 | 1370 | 1250 | 1530 |
| gas | - | 614 | 472 | 746 | 697 | 533 | *513* | *513* | 492 | **513** | 839 | 588 | 521 | 682 | 750 | 462 | 532 | *513* | *513* | *513* | 465 | 407 | *513* | 441 | 615 | *513* | *513* | 1090 | 694 |
|  | cogeneration | 529 | 503 | 936 | 840 | 351 | 455 | 423 | *475* | **475** | 530 | 671 | *475* | 173 | 648 | 173 | 496 | 629 | 599 | *475* | 450 | 523 | 542 | *475* | 686 | 810 | 555 | 436 | 652 |
| oil | - | 1160 | 913 | 1670 | 1060 | 877 | 1240 | 1180 | 866 | **1020** | 447 | 953 | 1320 | 993 | 1130 | 919 | 1060 | *1020* | *1020* | *1020* | *1020* | *1020* | *1020* | 834 | 1000 | *1020* | 854 | 1390 | 960 |
|  | cogeneration | 959 | 854 | 965 | 1520 | 680 | 965 | 873 | *935* | **935** | 952 | 770 | *935* | 1080 | 873 | *935* | 904 | 1530 | *935* | *935* | 1080 | *935* | 880 | 610 | 1260 | *935* | 837 | 873 | 1400 |
| low-voltage mix | - | 323 | 239 | 675 | 794 | 657 | 393 | 921 | 369 | **446** | 244 | 54.9 | 805 | 973 | 487 | 588 | 443 | 729 | 780 | 738 | 610 | 30.5 | 1030 | 400 | 474 | 940 | 42.3 | 447 | 458 |
| solar | - | 107 | 112 | 77.9 | 118 | 110 | *94.6* | *94.6* | 71.4 | **94.6** | *94.6* | 90.9 | *94.6* | 76.3 | *94.6* | *94.6* | 81.3 | *94.6* | *94.6* | *94.6* | 109 | *94.6* | *94.6* | 69.3 | 88.1 | 82.8 | 110 | 83.0 | 90.7 |

3

Table 2: CO$_2$ equivalent intensity per technology and country, embodied in infrastructure, in g CO$_2$ eq./kWh. Values in italic indicate that the country-specific factor is not available, and was replaced by the European weighted average for that technology (shown in bold).

| category | variant | AT | BE | BG | CZ | DE | DK | EE | ES | EU28 | FI | FR | GB | GR | HU | IE | IT | LT | LV | ME | NL | NO | PL | PT | RO | RS | SE | SI | SK |
|---|---|---|---|---|---|---|---|---|---|---|---|---|---|---|---|---|---|---|---|---|---|---|---|---|---|---|---|---|---|
| high-voltage mix | - | 5.48 | 3.10 | 3.10 | 2.39 | 4.49 | 7.41 | 3.64 | 5.23 | **4.04** | 3.16 | 3.02 | 1.18 | 3.60 | 2.63 | 4.17 | 6.18 | 6.17 | 3.80 | 4.04 | 2.40 | 6.55 | 2.82 | 6.66 | 5.42 | 3.36 | 4.41 | 2.32 | 2.56 |
| wind | - | 17.6 | 16.1 | 19.4 | 19.2 | 19.8 | 13.7 | 19.6 | 14.0 | **16.7** | 22.8 | 15.5 | 16.7 | 15.0 | 13.5 | 13.6 | 19.5 | 13.2 | 18.0 | 16.7 | 16.2 | 14.2 | 16.4 | 13.6 | 25.2 | 16.7 | 16.1 | 16.7 | 16.7 |
| nuclear | - | 2.10 | 1.93 | 1.93 | 1.93 | 1.89 | 2.10 | 2.10 | 1.95 | **2.10** | 1.99 | 2.27 | 2.10 | 2.10 | 1.93 | 2.10 | 2.10 | 2.10 | 2.10 | 2.10 | 1.93 | 2.10 | 2.10 | 2.10 | 1.86 | 2.10 | 1.96 | 1.93 | 1.93 |
| geothermal | - | 81.8 | 81.8 | 81.8 | 81.8 | 81.8 | 81.8 | 81.8 | 81.8 | **81.8** | 81.8 | 81.8 | 81.8 | 81.8 | 81.8 | 81.8 | 81.8 | 81.8 | 81.8 | 81.8 | 81.8 | 81.8 | 81.8 | 81.8 | 81.8 | 81.8 | 81.8 | 81.8 | 81.8 |
| biomass | cogeneration | 3.40 | 3.40 | 3.40 | 3.40 | 3.40 | 3.40 | 3.40 | 3.40 | **3.40** | 3.40 | 3.40 | 3.40 | 3.40 | 3.40 | 3.40 | 3.40 | 3.40 | 3.40 | 3.40 | 3.40 | 3.40 | 3.40 | 3.40 | 3.40 | 3.40 | 3.40 | 3.40 | 3.40 |
| hydropower | pumped storage | 6.52 | 6.52 | 6.52 | 6.52 | 6.52 | 6.52 | 6.52 | 6.52 | **6.52** | 6.52 | 6.52 | 6.52 | 6.52 | 6.52 | 6.52 | 6.52 | 6.52 | 6.52 | 6.52 | 6.52 | 6.52 | 6.52 | 6.52 | 6.52 | 6.52 | 6.52 | 6.52 | 6.52 |
| | reservoir | 6.52 | 6.52 | 6.52 | 6.52 | 6.52 | 6.52 | 6.52 | 6.52 | **6.52** | 6.52 | 6.52 | 6.52 | 6.52 | 6.52 | 6.52 | 6.52 | 6.52 | 6.52 | 6.52 | 6.52 | 6.52 | 6.52 | 6.52 | 6.52 | 6.52 | 6.52 | 6.52 | 6.52 |
| | run-of-river | 4.39 | 4.39 | 4.39 | 4.39 | 4.39 | 4.39 | 4.39 | 4.39 | **4.39** | 4.39 | 4.39 | 4.39 | 4.39 | 4.39 | 4.39 | 4.39 | 4.39 | 4.39 | 4.39 | 4.39 | 4.39 | 4.39 | 4.39 | 4.39 | 4.39 | 4.39 | 4.39 | 4.39 |
| coal | - | 1.37 | 1.58 | 2.41 | 2.46 | 2.13 | 1.96 | 2.34 | 1.69 | **1.96** | 1.87 | 1.58 | 1.55 | 2.38 | 3.01 | 1.84 | 1.57 | 2.40 | 1.96 | 1.96 | 1.46 | 1.96 | 1.96 | 1.56 | 2.45 | 2.82 | 2.40 | 2.59 | 1.96 |
| | cogeneration | 1.17 | 1.82 | 1.82 | 1.66 | 1.34 | 1.43 | 1.82 | 1.82 | **1.82** | 1.20 | 1.82 | 1.82 | 2.37 | 2.25 | 1.82 | 1.42 | 1.82 | 1.82 | 1.82 | 1.58 | 1.25 | 1.87 | 1.82 | 2.25 | 2.25 | 1.02 | 2.24 | 1.96 |
| gas | - | 0.916 | 0.927 | 0.469 | 0.559 | 0.818 | 0.721 | 0.721 | 1.15 | **0.721** | 1.82 | 1.25 | 0.354 | 1.08 | 0.962 | 0.880 | 0.964 | 0.721 | 0.721 | 0.721 | 0.809 | 0.704 | 0.721 | 1.15 | 0.386 | 0.721 | 0.721 | 0.688 | 0.878 |
| | cogeneration | 2.30 | 2.48 | 4.96 | 4.35 | 4.31 | 3.35 | 4.00 | 5.69 | **3.19** | 2.56 | 3.79 | 3.19 | 5.69 | 2.85 | 5.69 | 2.78 | 3.37 | 3.02 | 3.19 | 1.89 | 2.32 | 1.88 | 3.19 | 3.57 | 4.31 | 3.71 | 3.91 | 2.54 |
| oil | - | 2.64 | 2.08 | 3.75 | 2.39 | 1.99 | 2.76 | 2.65 | 1.94 | **2.27** | 1.02 | 2.14 | 2.97 | 2.19 | 2.60 | 2.06 | 2.32 | 2.27 | 2.27 | 2.27 | 2.27 | 2.27 | 2.27 | 1.89 | 2.24 | 2.27 | 1.95 | 3.12 | 2.16 |
| | cogeneration | 2.19 | 1.94 | 2.17 | 3.42 | 1.54 | 2.15 | 1.96 | 2.08 | **2.08** | 2.18 | 1.73 | 2.08 | 2.37 | 1.96 | 2.08 | 1.98 | 3.42 | 2.08 | 2.08 | 2.53 | 2.08 | 1.98 | 1.39 | 2.89 | 2.08 | 1.91 | 1.96 | 3.14 |
| low-voltage mix | - | 4.54 | 2.99 | 6.18 | 7.71 | 13.8 | 2.97 | 2.95 | 6.76 | **6.41** | 2.99 | 3.66 | 2.96 | 12.1 | 2.95 | 2.96 | 13.0 | 4.02 | 2.97 | 2.91 | 3.01 | 2.97 | 2.97 | 3.68 | 5.88 | 2.93 | 3.03 | 3.03 | 3.44 |
| solar | - | 107 | 112 | 77.9 | 118 | 110 | 94.6 | 94.6 | 71.4 | **94.6** | 94.6 | 90.9 | 94.6 | 76.3 | 94.6 | 94.6 | 81.3 | 109 | 94.6 | 94.6 | 109 | 94.6 | 94.6 | 69.2 | 88.1 | 82.7 | 110 | 83.0 | 90.7 |

4

Table 3: CO$_2$ equivalent intensity per technology and country, embodied in operations, in g CO$_2$ eq./kWh. Values in italic indicate that the country-specific factor is not available, and was replaced by the European weighted average for that technology (shown in bold).

| category | variant | AT | BE | BG | CZ | DE | DK | EE | ES | EU28 | FI | FR | GB | GR | HU | IE | IT | LT | LV | ME | NL | NO | PL | PT | RO | RS | SE | SI | SK |
|---|---|---|---|---|---|---|---|---|---|---|---|---|---|---|---|---|---|---|---|---|---|---|---|---|---|---|---|---|---|
| high-voltage mix | - | 119 | 185 | 606 | 729 | 649 | 425 | 1030 | 331 | **422** | 259 | 38.8 | 800 | 976 | 397 | 509 | 463 | 545 | 516 | 422 | 614 | 9.38 | 998 | 354 | 392 | 848 | 17.4 | 432 | 214 |
| wind | - | 0.149 | 0.156 | 0.149 | 0.166 | 0.165 | 0.126 | 0.165 | 0.122 | **0.142** | 0.156 | 0.133 | 0.142 | 0.121 | 0.114 | 0.116 | 0.161 | 0.110 | 0.159 | *0.142* | 0.133 | 0.120 | 0.140 | 0.117 | 0.192 | *0.142* | 0.141 | *0.142* | *0.142* |
| nuclear | - | *10.3* | 10.1 | 10.1 | 10.1 | 9.37 | *10.3* | *10.3* | 10.2 | **10.3** | 10.5 | 10.6 | *10.3* | *10.3* | 10.1 | *10.3* | *10.3* | *10.3* | *10.3* | *10.3* | 10.1 | *10.3* | *10.3* | *10.3* | 12.3 | *10.3* | 10.3 | 10.1 | 10.1 |
| geothermal | - | *0.00664* | *0.00664* | *0.00664* | *0.00664* | *0.00664* | *0.00664* | *0.00664* | *0.00664* | **0.00664** | *0.00664* | *0.00664* | *0.00664* | *0.00664* | *0.00664* | *0.00664* | *0.00664* | *0.00664* | *0.00664* | *0.00664* | *0.00664* | *0.00664* | *0.00664* | *0.00664* | *0.00664* | *0.00664* | *0.00664* | *0.00664* | *0.00664* |
| biomass | cogeneration | 50.4 | 50.4 | 53.4 | 50.4 | 50.4 | 50.4 | 53.4 | 50.4 | **50.5** | 50.4 | 50.4 | 50.4 | 50.5 | 50.4 | 50.4 | 50.4 | 53.4 | 53.4 | *50.5* | 50.4 | 50.5 | 50.4 | 50.4 | 50.4 | 50.5 | 50.4 | 50.4 | 50.4 |
| hydropower | pumped storage | 445 | 372 | 894 | 1140 | 958 | 611 | 611 | 539 | **611** | 611 | 70.8 | 611 | 1410 | 611 | 845 | 608 | 1030 | 611 | *611* | 611 | 35.2 | 1410 | 582 | 622 | 1210 | 611 | 610 | 678 |
|  | reservoir | 0.445 | 8.13 | 8.13 | 44.8 | 44.8 | 8.13 | 8.13 | 44.8 | **8.13** | 44.8 | 0.445 | 8.13 | 8.13 | 8.13 | 8.13 | 0.445 | 8.13 | 8.13 | *8.13* | 8.13 | 0.445 | 8.13 | 44.8 | 8.13 | 0.445 | 44.8 | 8.13 | 44.8 |
|  | run-of-river | 0.0253 | 0.0253 | 0.0253 | 0.0253 | 0.0253 | 0.0253 | 0.0253 | 0.0253 | **0.0253** | 0.0253 | 0.0253 | 0.0253 | 0.0253 | 0.0253 | 0.0253 | 0.0253 | 0.0253 | 0.0253 | *0.0253* | 0.0253 | 0.0253 | 0.0253 | 0.0253 | 0.0253 | 0.0253 | 0.0253 | 0.0253 | 0.0253 |
| coal | - | 984 | 1120 | 1180 | 1180 | 1160 | 1160 | 1300 | 1210 | **1160** | 1080 | 1090 | 1140 | 1300 | 1400 | 1070 | 1150 | 1170 | 1160 | *1160* | 1030 | 1160 | 1160 | 1140 | 1140 | 1340 | 1170 | 1190 | 1160 |
|  | cogeneration | 1220 | 1210 | 1250 | 1710 | 1160 | 1050 | 1210 | 1210 | **1210** | 1100 | 1210 | 1210 | 1560 | 1230 | 1210 | 1260 | 1210 | 1210 | *1210* | 996 | 1490 | 1160 | 1210 | 1230 | 1230 | 1370 | 1240 | 1530 |
| gas | - | 613 | 471 | 745 | 696 | 533 | 513 | 513 | 491 | **513** | 837 | 587 | 521 | 681 | 749 | 461 | 531 | 513 | 513 | *513* | 464 | 406 | 513 | 440 | 615 | 513 | 513 | 1090 | 694 |
|  | cogeneration | 527 | 501 | 932 | 835 | 347 | 452 | 419 | 167 | **471** | 528 | 668 | 471 | 167 | 645 | 167 | 493 | 625 | 596 | *471* | 449 | 520 | 540 | 471 | 682 | 805 | 551 | 432 | 649 |
| oil | - | 1150 | 911 | 1660 | 1060 | 875 | 1240 | 1180 | 864 | **1010** | 446 | 951 | 1320 | 990 | 1130 | 917 | 1060 | 1010 | 1010 | *1010* | 1010 | 1010 | 1010 | 832 | 997 | 1010 | 852 | 1380 | 958 |
|  | cogeneration | 957 | 852 | 962 | 1520 | 678 | 963 | 871 | 933 | **933** | 949 | 768 | 933 | 1070 | 871 | 933 | 902 | 1530 | 933 | *933* | 1070 | 933 | 878 | 609 | 1250 | 933 | 835 | 871 | 1390 |
| low-voltage mix | - | 319 | 236 | 669 | 786 | 643 | 390 | 918 | 362 | **440** | 241 | 51.2 | 802 | 961 | 484 | 585 | 430 | 725 | 777 | 735 | 607 | 27.5 | 1030 | 396 | 468 | 937 | 39.3 | 444 | 455 |
| solar | - | 0.00580 | 0.00502 | 0.00423 | 0.00642 | 0.00448 | 0.00349 | 0.00349 | 0.00234 | **0.00349** | 0.00349 | 0.00370 | 0.00349 | 0.00415 | 0.00349 | 0.00349 | 0.00166 | 0.00591 | 0.00349 | 0.00349 | 0.00591 | 0.00349 | 0.00349 | 0.00185 | 0.00478 | 0.00445 | 0.00565 | 0.00453 | 0.00493 |

5

# B  Flow tracing

## B.1  Formulation

**Nomenclature**

| | |
|---|---|
| $\alpha$ | set of all generation/storage technologies. |
| $L_n$ | nodal load. |
| $F_{n \to k}$ | nodal outflow to direct neighbors. |
| $F_{m \to n}$ | nodal inflow from direct neighbors. |
| $G_{n,\alpha}$ | nodal generation for all technologies. |
| $S^+_{n,\alpha}$ | storage discharge for each storage technology $\alpha$ at node $n$. |
| $S^-_n$ | sum of storage charging at node $n$. |
| $q_{n,\alpha}$ | nodal colormix. |

The nodal color mix refers to the mixing of electricity at each node from different technologies and countries of origin, where each technology for each country has been assigned a unique color [2]. Note that this is an assumption, analogous to the mixing of water flows in pipes, used to approximate the mixing of power flows at nodes in the transmission system.

Figure 1 shows a sketch of the flow tracing implementation. For every hour all imports, generation, and storage discharge are mixed equally in the node, which then determines the color mix of the exports and the power serving the local load. We do not keep track of the color mix flowing into storage, but track which storage type the power originated from when the storages are discharging. This mixing approach is called *average participation* or *proportional sharing* in the literature which was also proposed initially in [3]. For a discussion of different allocation methods, see [4]. For comprehensive reviews, see [5, 6].

The sketch in Figure 1 describes the nodal power balance

$$L_n + S^-_n + \sum_k F_{n \to k} = \sum_\alpha \left( G_{n,\alpha} + S^+_{n,\alpha} \right) + \sum_m F_{m \to n}, \tag{1}$$

where the left-hand side and the right-hand side account for the flows out of and into a node, respectively. In this, and following equations, there is an implicit time index as the flow tracing is performed for every hour. We include nodal color mixes in the nodal power balance

$$q_{n,\alpha} \left( L_n + S^-_n + \sum_k F_{n \to k} \right) = G_{n,\alpha} + S^+_{n,\alpha} + \sum_m q_{m,\alpha} F_{m \to n}, \tag{2}$$

which is now an equation per country $n$ per technology type $\alpha$. Rearranging (2) we can write a matrix formula describing a unique solution for the nodal power mix $q_{n,\alpha}$ according to [7]:

$$\sum_m \left[ \delta_{n,m} \left( L_m + S^-_m + \sum_k F_{m \to k} \right) - F_{m \to n} \right] q_{m,\alpha} = G_{n,\alpha} + S^+_{n,\alpha}. \tag{3}$$

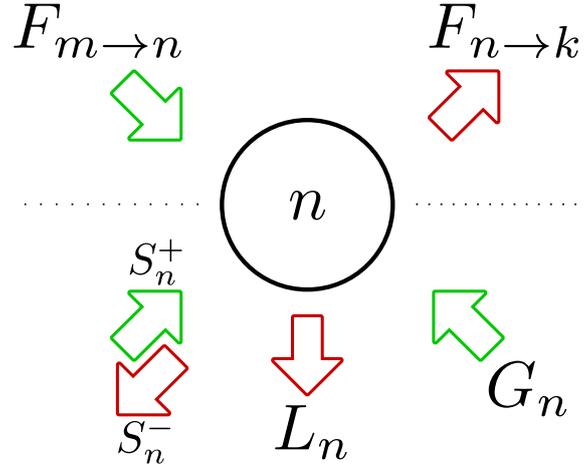

Figure 1: Sketch of flow tracing methodology.

Here $q_{m,\alpha}$ is the hourly nodal color mix for node $m$ split into components for every technology for every country. The $\alpha$ set allows us to track originating technology as well as originating country e.g. we can trace who is consuming Danish wind power. Multiplying the nodal color mix with the nodal load and the carbon intensity of the originating generation/storage technologies allows us to calculate consumption-based carbon intensity allocation.

## B.2 Handling of missing data

As we are using raw data directly from the power system there will be occurrences of missing values. In case of missing data for production or imports/exports for a country the particular country is excluded from the flow tracing calculation for that specific hour.

Imports from countries not included in the topology are included (e.g. Switzerland), but do not have an effect on the nodal mix of the importer (they simply scale the color mix, but do not change the ratios). Exports to countries outside the considered topology are subtracted.

Figure 2 shows $\sum_\alpha q_{n,\alpha}$ for every country for every hour. If (3) is perfectly balanced it should be the case that $\sum_\alpha q_{n,\alpha} = 1$. Cases of partially missing data leads to $\sum_\alpha q_{n,\alpha} \neq 1$. This is usually caused by one country being excluded due to missing data (which explains the occurrence of 0's in Figure 2), which affects the nodal balance of neighboring countries. See e.g. the effect of missing data for Ireland on Great Britain. We observe no cases of $\sum_\alpha q_{n,\alpha} > 1$. The missing data mostly occurs for small, satellite countries e.g. Ireland and Montenegro, which only have a small effect on the closest neighbors.

The total number of entries in Figure 2:

$$hours \cdot nodes = 8760 \cdot 28 = 245280 \tag{4}$$

Of these there are 6367 occurrences of $q_{n,\alpha} = 0$ (due to missing data), which is only 2.6%. When the occurrences of 0 are subtracted there are 3742 occurrences where $q_{n,\alpha} < .9999$ which is only 1.5%. The cases where $0 < q_{n,\alpha} < .9999$ are all rather close to 1 (all except 3

7

are above .8 and most are above .9). The occurrences of 0 are predominantly for Ireland, Montenegro and Estonia, which are both small countries at the edge of the network.

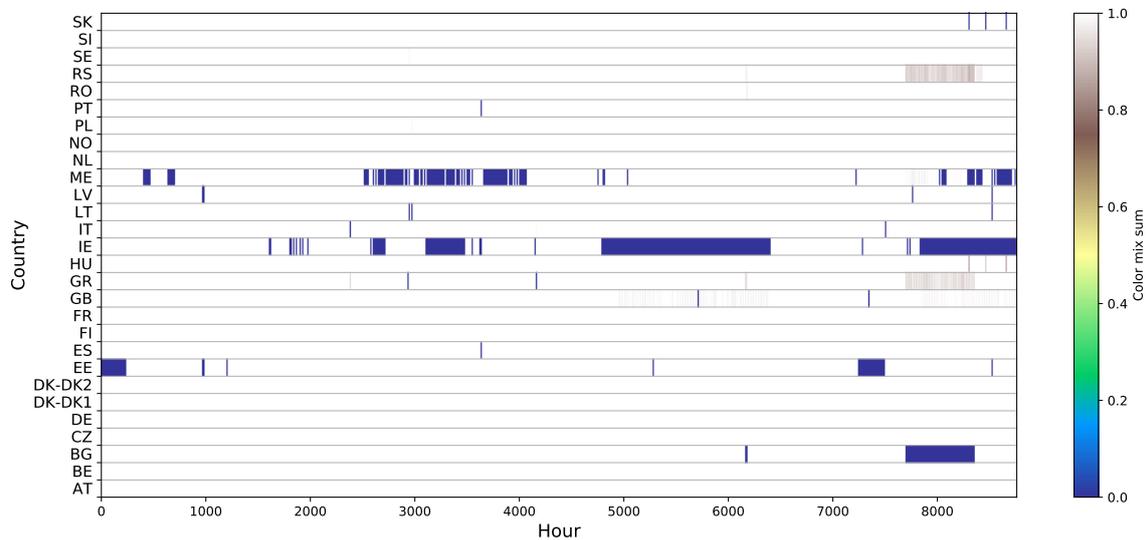

Figure 2: Flow tracing consistency check. Dark blue means generation or import/export data is entirely missing for a country, lighter colors mean it is partially complete, and white means fully complete data.

## C  Additional results

Figure 3 shows a comparison of hourly production intensity with hourly load for the full year of 2017 for every country. The production intensity is calculated based on the production within each country. The figure is split in two parts with large countries in the top panel and smaller countries in the bottom panel. In the top panel we see that Norway, Sweden and France have low intensities regardless of the level of consumption, which is due to a high share of hydro power in the Nordic countries and nuclear power in France. On the other hand, Poland has very high intensity due to a high share of coal power generation.

Figure 4 shows the stacked average consumption intensity per kWh per hour in Austria for all of 2017. This figure does not tell anything about the amount of power being consumed by each technology.

Figure 5 shows the total annual consumption intensity for Austria for 2017 based on flow tracing. From this figure we see that hydro is the technology providing most of the consumed power, but that the intensity from this consumption is among the lowest of the technologies. On the other hand coal power is one of the smaller contributors to the consumed power, but has the largest intensity.

Figure 6 shows average hourly production/consumption carbon intensity plotted as duration curves for Austria and Denmark e.g. if a country runs on 100% coal the entire year the

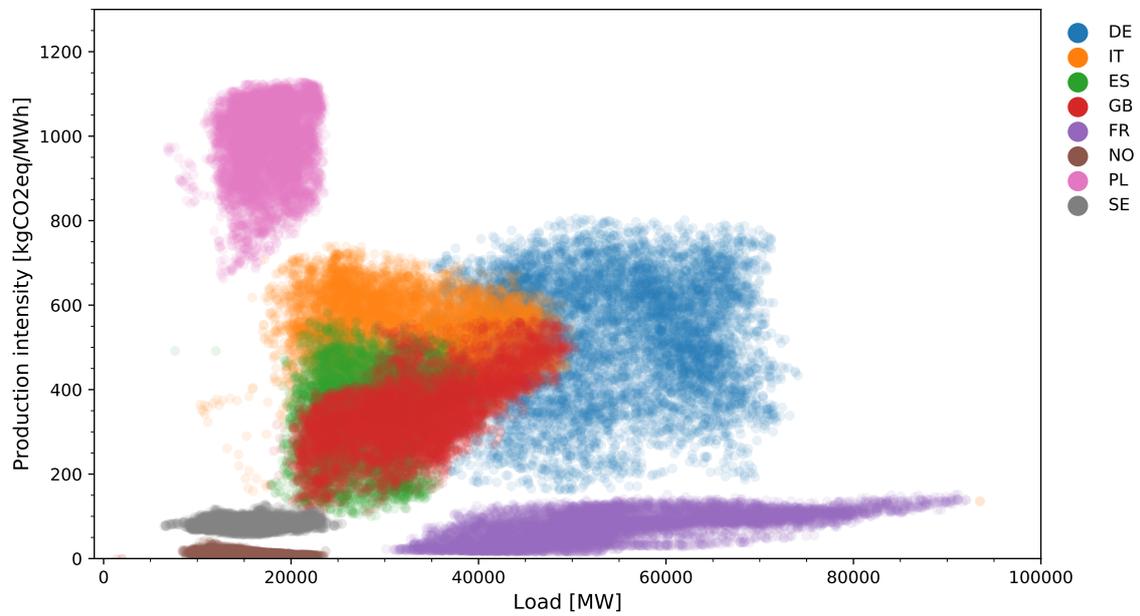
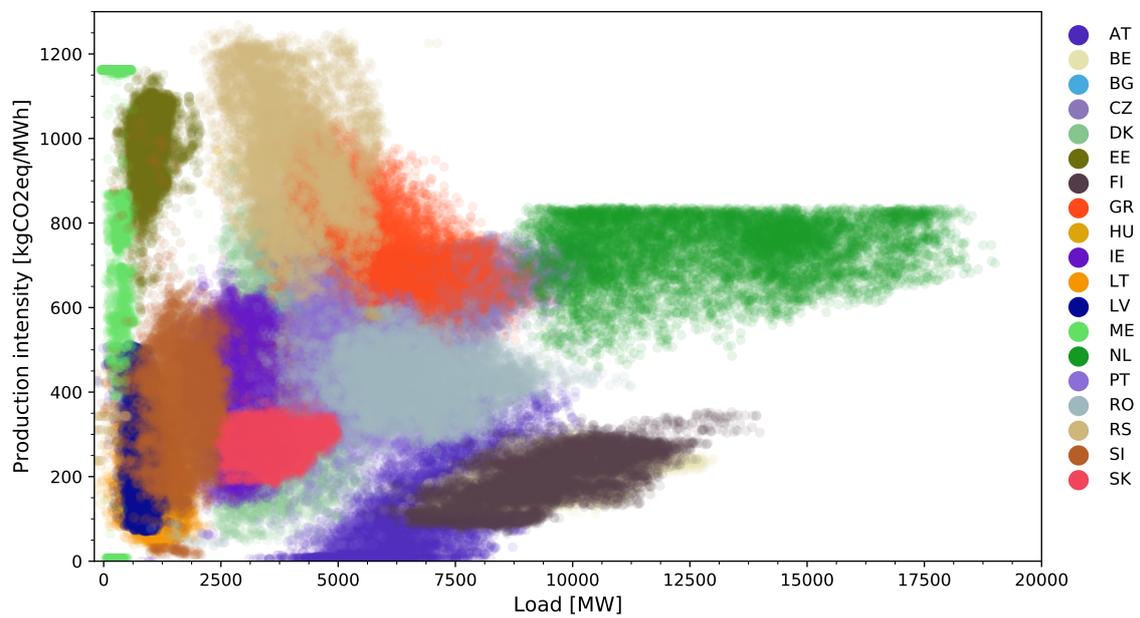

Figure 3: Comparison of hourly production intensity with hourly load for every country.

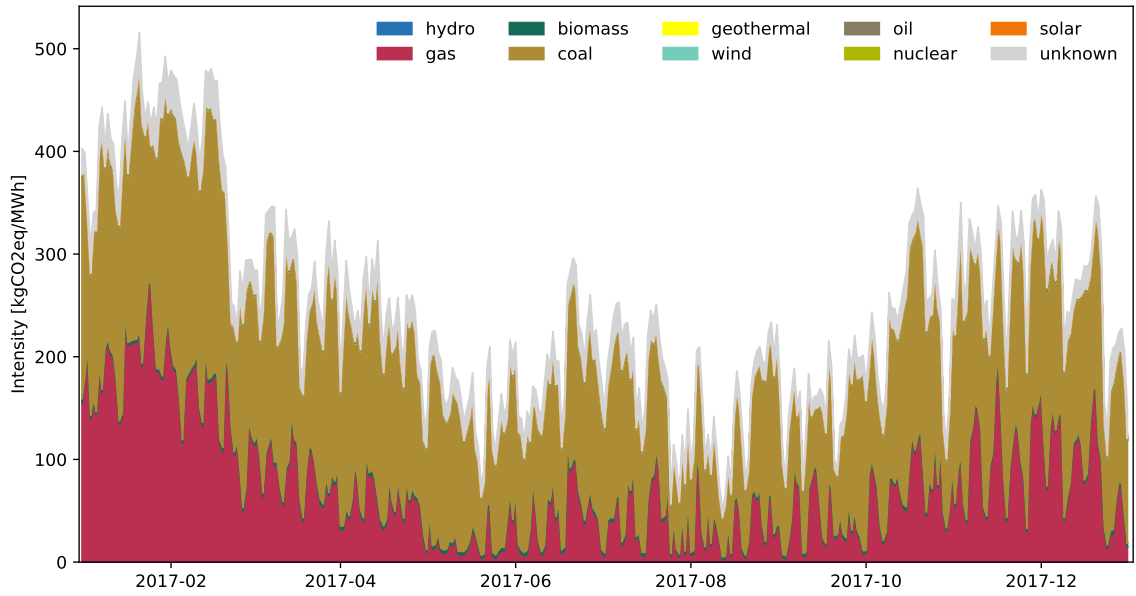

Figure 4: Hourly intensity per consumed unit of energy for Austria downsampled to daily averages.

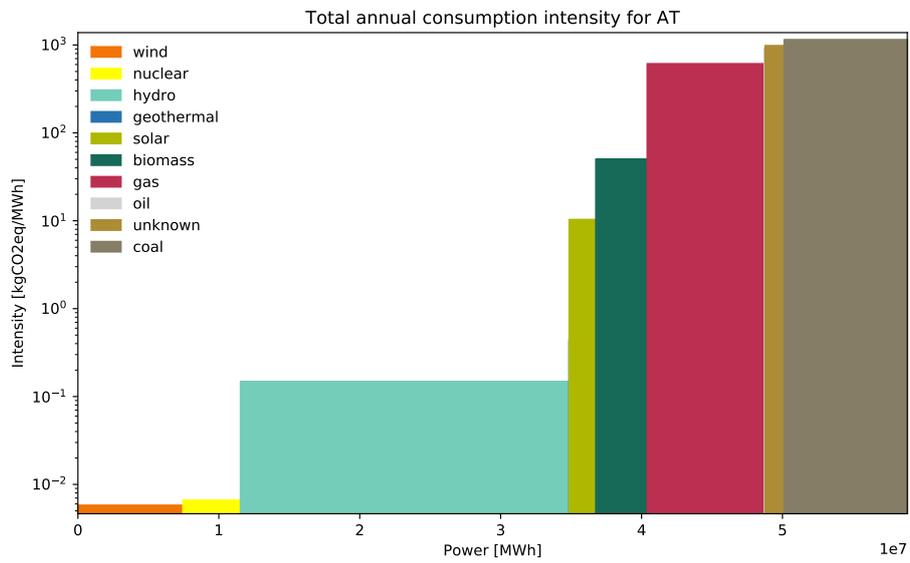

Figure 5: Total annual consumption intensity for Austria for 2017.

duration curve would be flat at that country's operational intensity for coal as seen in Table 3. This figure shows that AT has a low production intensity, but a higher consumption intensity due to imports. DK is relying on imports for a low consumption intensity since it has a high production intensity for approximately half of the year.

Figure 7 shows a comparison of average production (blue) and consumption (orange) intensity for each country. White dots mark the mean. The colored bars indicate 25%–75% quantiles and the gray bars 5%–95% quantiles. This is a summary of the duration curves for individual countries as shown in Figure 6.

Figure 8 shows the difference between production and consumption intensity as function of the share of non-fossil production of total production. Size of circles are proportional to average production. A value above zero corresponds to the country having a higher consumption intensity than production intensity. The figure shows a general trend that the higher the share of non-fossil production the higher the consumption intensity is compared to the production intensity. This can be explained by countries with high share of non-fossil production tend to import from countries with lower share of non-fossil production which results in the importing country's consumption intensity being higher than its production intensity.

Table 4 shows average production and consumption intensity per country. These values are plotted in Figure 3 in the article, they are also shown as the white markers in Figure 7, and the difference for each country is shown in Figure 8.

Figure 9 shows average intensity per imported/exported unit of energy. When calculating the average imported/exported intensity between two countries only hours with actual transfers have been used. A white entry means no data and only occurs for ME and RS. The figure should be read as NO exporting mostly low intensity hydro to all countries whereas EE and PL are exporting oil and coal to all countries. This figure doesn't say anything about the amount of energy being transferred e.g. most of the column for ME is based on data for very few hours as ME is a small, poorly connected country. The values in Figure 9 are also shown in Table 5.

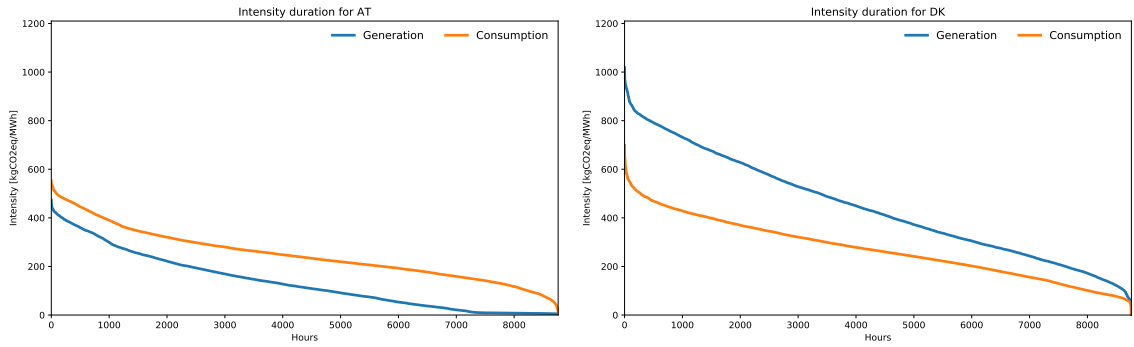

Figure 6: Average hourly production/consumption carbon intensity duration curves for Austria and Denmark.

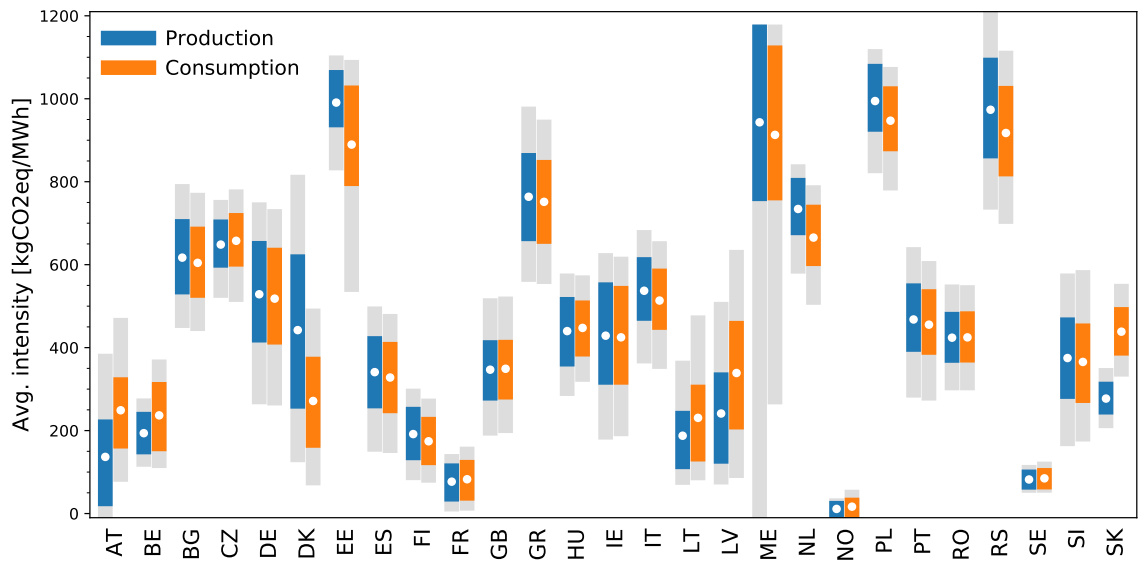

Figure 7: Comparison of average production (blue) and consumption (orange) intensity. White dots mark the mean, colored bars 25%-75% quantiles and the gray bars 5%-95% quantiles.

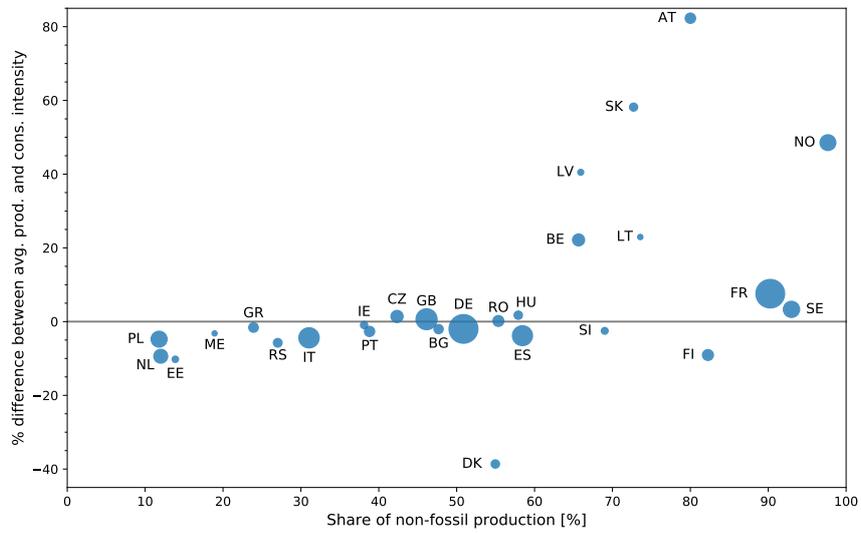

Figure 8: Difference between production and consumption intensity as function of the share of non-fossil production of total production. Size of circles are proportional to average production.

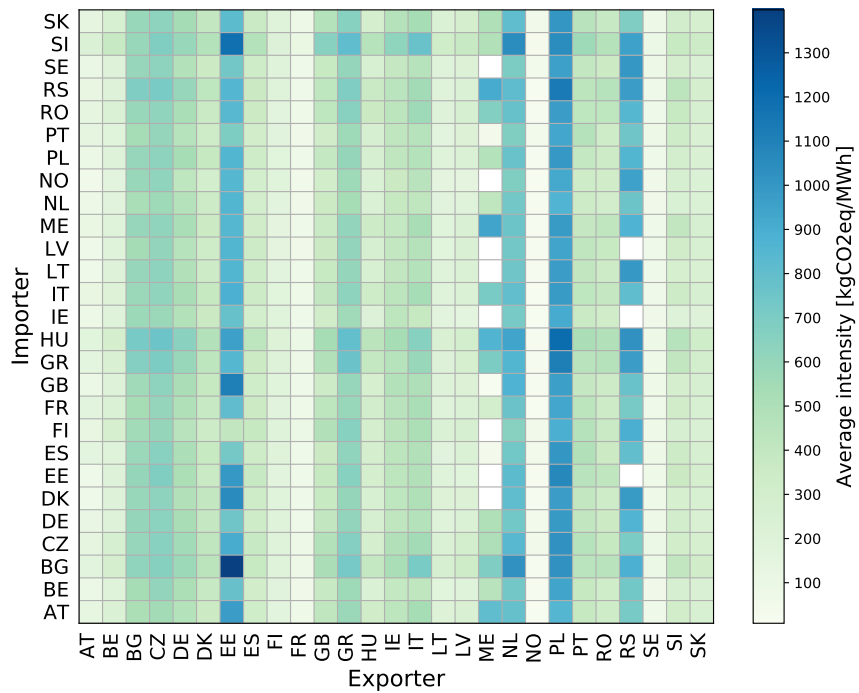

Figure 9: Average imported/exported intensity. White cells indicating missing data. This figure doesn't say anything about the amount of energy being transferred e.g. most of the column for ME is based on data for very few hours as ME is a small, poorly connected country.

13

Table 4: Average production and consumption intensity for each country. These values are plotted in Figure 3 in the article. Units are kgCO2eq/MWh.

| | AT | BE | BG | CZ | DE | DK | EE | ES | FI | FR | GB | GR | HU | IE | IT | LT | LV | ME | NL | NO | PL | PT | RO | RS | SE | SI | SK |
|---|---|---|---|---|---|---|---|---|---|---|---|---|---|---|---|---|---|---|---|---|---|---|---|---|---|---|---|
| Consumption | 248 | 236 | 604 | 657 | 518 | 271 | 889 | 327 | 174 | 82 | 349 | 751 | 447 | 424 | 513 | 230 | 338 | 912 | 665 | 16 | 947 | 455 | 424 | 917 | 84 | 365 | 438 |
| Production | 136 | 193 | 616 | 648 | 528 | 442 | 990 | 340 | 191 | 76 | 346 | 763 | 439 | 428 | 537 | 187 | 241 | 943 | 734 | 11 | 994 | 467 | 424 | 973 | 82 | 374 | 277 |

Table 5: Average intensity of power imported and exported between countries. These values are plotted in Figure 9. Columns are exporters and rows are importers. Units are kgCO2eq/MWh.

|    | AT  | BE  | BG  | CZ  | DE  | DK  | EE   | ES  | FI  | FR  | GB  | GR  | HU  | IE  | IT  | LT  | LV  | ME  | NL   | NO | PL   | PT  | RO  | RS   | SE | SI  | SK  |
|----|-----|-----|-----|-----|-----|-----|------|-----|-----|-----|-----|-----|-----|-----|-----|-----|-----|-----|------|----|------|-----|-----|------|----|-----|-----|
| AT | 136 | 232 | 503 | 546 | 469 | 360 | 974  | 322 | 160 | 73  | 417 | 580 | 325 | 428 | 533 | 211 | 245 | 808 | 787  | 15 | 853  | 385 | 344 | 714  | 68 | 316 | 255 |
| BE | 91  | 193 | 533 | 615 | 506 | 381 | 774  | 316 | 181 | 65  | 377 | 572 | 227 | 394 | 417 | 174 | 195 | 452 | 734  | 19 | 938  | 384 | 322 | 736  | 67 | 265 | 237 |
| BG | 164 | 281 | 616 | 662 | 572 | 438 | 1397 | 372 | 196 | 85  | 513 | 719 | 395 | 518 | 708 | 273 | 342 | 690 | 1019 | 27 | 1021 | 447 | 450 | 891  | 91 | 406 | 324 |
| CZ | 143 | 258 | 599 | 648 | 552 | 413 | 904  | 377 | 196 | 86  | 479 | 668 | 285 | 480 | 549 | 234 | 294 | 501 | 843  | 59 | 1021 | 454 | 386 | 704  | 90 | 312 | 274 |
| DE | 121 | 206 | 599 | 640 | 528 | 391 | 748  | 336 | 192 | 73  | 406 | 613 | 263 | 412 | 470 | 193 | 234 | 493 | 739  | 55 | 989  | 406 | 360 | 869  | 82 | 302 | 244 |
| DK | 71  | 206 | 590 | 631 | 489 | 402 | 1050 | 335 | 182 | 69  | 375 | 600 | 256 | 420 | 491 | 210 | 221 | -   | 808  | 20 | 969  | 399 | 343 | 981  | 75 | 275 | 256 |
| EE | 70  | 200 | 601 | 681 | 505 | 370 | 990  | 381 | 170 | 69  | 383 | 656 | 286 | 430 | 490 | 191 | 182 | -   | 817  | 21 | 1066 | 451 | 413 | -    | 76 | 363 | 281 |
| ES | 149 | 193 | 582 | 651 | 508 | 374 | 727  | 340 | 191 | 68  | 365 | 617 | 287 | 433 | 481 | 179 | 226 | 44  | 737  | 55 | 1003 | 484 | 359 | 795  | 83 | 351 | 260 |
| FI | 126 | 279 | 552 | 599 | 448 | 347 | 408  | 382 | 191 | 90  | 483 | 655 | 284 | 437 | 503 | 227 | 262 | -   | 649  | 62 | 892  | 467 | 369 | 894  | 77 | 268 | 237 |
| FR | 167 | 231 | 533 | 610 | 496 | 382 | 804  | 356 | 181 | 76  | 429 | 594 | 307 | 450 | 518 | 193 | 235 | 292 | 766  | 14 | 925  | 432 | 348 | 715  | 68 | 327 | 256 |
| GB | 94  | 203 | 556 | 628 | 517 | 387 | 1106 | 322 | 183 | 69  | 346 | 594 | 277 | 477 | 501 | 208 | 221 | 8   | 871  | 21 | 965  | 389 | 345 | 772  | 77 | 283 | 272 |
| GR | 149 | 246 | 668 | 698 | 591 | 429 | 847  | 359 | 199 | 80  | 487 | 763 | 394 | 462 | 593 | 219 | 276 | 698 | 859  | 24 | 1113 | 448 | 469 | 976  | 75 | 409 | 293 |
| HU | 174 | 280 | 717 | 758 | 644 | 475 | 961  | 427 | 212 | 90  | 534 | 792 | 439 | 529 | 656 | 244 | 297 | 861 | 939  | 25 | 1206 | 515 | 489 | 1022 | 83 | 457 | 317 |
| IE | 86  | 197 | 577 | 581 | 480 | 361 | 782  | 297 | 185 | 70  | 326 | 561 | 214 | 427 | 386 | 169 | 154 | -   | 717  | 18 | 913  | 367 | 326 | -    | 63 | 205 | 203 |
| IT | 126 | 214 | 579 | 617 | 526 | 387 | 892  | 350 | 170 | 73  | 405 | 630 | 336 | 432 | 537 | 192 | 229 | 710 | 797  | 13 | 983  | 423 | 392 | 806  | 68 | 349 | 263 |
| LT | 68  | 199 | 596 | 630 | 486 | 354 | 856  | 337 | 165 | 68  | 371 | 605 | 255 | 406 | 469 | 187 | 235 | -   | 750  | 20 | 973  | 405 | 346 | 992  | 70 | 274 | 244 |
| LV | 69  | 194 | 540 | 615 | 456 | 329 | 854  | 346 | 156 | 67  | 363 | 610 | 267 | 403 | 468 | 173 | 240 | -   | 733  | 21 | 944  | 412 | 375 | -    | 69 | 326 | 255 |
| ME | 114 | 194 | 594 | 618 | 522 | 359 | 848  | 325 | 158 | 69  | 371 | 613 | 349 | 396 | 526 | 183 | 241 | 942 | 759  | 22 | 984  | 388 | 413 | 872  | 70 | 408 | 262 |
| NL | 85  | 177 | 523 | 567 | 466 | 348 | 855  | 294 | 166 | 62  | 356 | 531 | 233 | 386 | 416 | 177 | 193 | 416 | 734  | 19 | 868  | 353 | 308 | 759  | 66 | 247 | 226 |
| NO | 57  | 167 | 589 | 620 | 422 | 298 | 845  | 291 | 183 | 59  | 313 | 569 | 234 | 364 | 448 | 159 | 159 | -   | 684  | 11 | 920  | 341 | 327 | 957  | 66 | 261 | 231 |
| PL | 94  | 203 | 595 | 635 | 529 | 385 | 857  | 330 | 171 | 69  | 396 | 604 | 262 | 417 | 473 | 188 | 225 | 470 | 770  | 20 | 994  | 396 | 348 | 858  | 70 | 291 | 244 |
| PT | 145 | 182 | 537 | 606 | 463 | 334 | 693  | 314 | 177 | 66  | 327 | 565 | 268 | 414 | 458 | 166 | 221 | 44  | 683  | 55 | 926  | 467 | 333 | 746  | 82 | 329 | 245 |
| RO | 138 | 229 | 593 | 616 | 520 | 370 | 842  | 356 | 170 | 76  | 424 | 652 | 363 | 431 | 565 | 201 | 258 | 673 | 778  | 23 | 971  | 425 | 424 | 849  | 72 | 376 | 268 |
| RS | 127 | 206 | 691 | 709 | 596 | 420 | 853  | 363 | 189 | 71  | 425 | 686 | 367 | 444 | 549 | 187 | 234 | 913 | 815  | 13 | 1134 | 436 | 458 | 973  | 72 | 435 | 283 |
| SE | 97  | 198 | 599 | 631 | 477 | 357 | 731  | 335 | 191 | 71  | 372 | 608 | 254 | 389 | 454 | 191 | 224 | -   | 701  | 55 | 961  | 404 | 355 | 997  | 82 | 282 | 241 |
| SI | 226 | 379 | 592 | 685 | 587 | 470 | 1184 | 472 | 217 | 111 | 652 | 805 | 463 | 622 | 772 | 322 | 379 | 486 | 1049 | 33 | 1048 | 565 | 466 | 955  | 92 | 374 | 347 |
| SK | 140 | 254 | 588 | 635 | 540 | 402 | 820  | 371 | 191 | 85  | 474 | 658 | 280 | 467 | 533 | 226 | 286 | 492 | 806  | 59 | 1002 | 449 | 380 | 690  | 87 | 305 | 277 |



\end{document}